\documentclass[twoside,a4paper]{article}
\usepackage{etoolbox}
\usepackage{preamble/dafx_25}
\usepackage{amsmath,amssymb,amsfonts,amsthm}
\usepackage{euscript}
\usepackage[T1]{fontenc}
\usepackage[utf8]{inputenc}
\usepackage{ifpdf}
\usepackage[english]{babel}
\usepackage{caption}
\usepackage{subfig} 
\usepackage{color}
\usepackage{amsmath, amsfonts,amssymb}
\usepackage{tabularx}
\usepackage{algorithm}    
\usepackage{algpseudocode}
\usepackage{arydshln}
\usepackage{booktabs} 

\input glyphtounicode
\ninept

\newcounter{numauth}
\newcounter{listcnt}
\newcommand\authcnt[1]{\ifdefined#1 \stepcounter{numauth} \fi}

\newcommand\addauth[1]{
\ifdefined#1 
\stepcounter{listcnt}
\ifnum \value{listcnt}<\value{numauth}
\appto\authorslist{, #1}
\else
\appto\authorslist{~and~#1}
\fi
\fi}
\authcnt{\paperauthorB}
\authcnt{\paperauthorC}
\authcnt{\paperauthorD}
\authcnt{\paperauthorE}
\authcnt{\paperauthorF}
\authcnt{\paperauthorG}
\authcnt{\paperauthorH}
\authcnt{\paperauthorI}
\authcnt{\paperauthorJ}
\def\authorslist{\paperauthorA}
\addauth{\paperauthorB}
\addauth{\paperauthorC}
\addauth{\paperauthorD}
\addauth{\paperauthorE}
\addauth{\paperauthorF}
\addauth{\paperauthorG}
\addauth{\paperauthorH}
\addauth{\paperauthorI}
\addauth{\paperauthorJ}

\usepackage{times}

\newif\ifpdf
\ifx\pdfoutput\relax
\else
   \ifcase\pdfoutput
      \pdffalse
   \else
      \pdftrue
   \fi
\fi

\def\papertitle{A statistics-driven differentiable approach for sound texture synthesis and analysis}
\def\paperauthorA{Esteban Gutiérrez}
\def\paperauthorB{Frederic Font}
\def\paperauthorC{Xavier Serra}
\def\paperauthorD{Lonce Wyse}

\ifpdf 
  \usepackage[pdftex,
    pdftitle={\papertitle},
    pdfauthor={\authorslist},
    pdfsubject={Proceedings of the 28th International Conference on Digital Audio Effects (DAFx25)},
    colorlinks=false, 
    bookmarksnumbered, 
    pdfstartview=XYZ 
  ]{hyperref}
  \usepackage[pdftex]{graphicx}
\else 
  \usepackage[dvips]{epsfig,graphicx}
  \usepackage[dvips,
    pdftitle={\papertitle},
    pdfauthor={\authorslist},
    pdfsubject={Proceedings of the 28th International Conference on Digital Audio Effects (DAFx25)},
    colorlinks=false, 
    bookmarksnumbered, 
    pdfstartview=XYZ 
  ]{hyperref}
\fi
\usepackage[hypcap=true]{caption}

\ifpdf 
  \DeclareGraphicsExtensions{.png,.jpg,.pdf}
\else  
  \DeclareGraphicsExtensions{.eps}
\fi
\def\papertitle{A statistics-driven differentiable approach for sound texture synthesis and analysis}
\def\paperauthorA{Esteban Gutiérrez}
\def\paperauthorB{Frederic Font}
\def\paperauthorC{Xavier Serra}
\def\paperauthorD{Lonce Wyse}

\title{\papertitle}
 \fouraffiliations{
\paperauthorA \,\thanks{\vspace{-3mm}}} {\href{https://www.upf.edu/web/mtg}{Music Technology Group} \\ Universitat Pompeu Fabra\\ Barcelona, Spain\\
 {\tt \href{mailto:esteban.gutierrezc@upf.edu}{esteban.gutierrezc@upf.edu}}
 }
 {\paperauthorB \,}
 {\href{https://www.upf.edu/web/mtg}{Music Technology Group} \\ Universitat Pompeu Fabra\\ Barcelona, Spain\\
 {\tt \href{mailto:frederic.font@upf.edu}{frederic.font@upf.edu}}
 }
 {\paperauthorC \,}
 {\href{https://www.upf.edu/web/mtg}{Music Technology Group} \\ Universitat Pompeu Fabra\\ Barcelona, Spain\\
 {\tt \href{mailto:xavier.serra@upf.edu}{xavier.serra@upf.edu}}
 }
 {\paperauthorD \,}
 {\href{https://www.upf.edu/web/mtg}{Music Technology Group} \\ Universitat Pompeu Fabra\\ Barcelona, Spain\\
 {\tt \href{mailto:lonce.wyse@upf.edu}{ lonce.wyse@upf.edu}}
 }

\usepackage{xcolor}

\begin{document}

\maketitle
\begin{abstract}
In this work, we introduce \texttt{TexStat}, a novel loss function specifically designed for the analysis and synthesis of texture sounds characterized by stochastic structure and perceptual stationarity. Drawing inspiration from the statistical and perceptual framework of McDermott and Simoncelli, \texttt{TexStat} identifies similarities between signals belonging to the same texture category without relying on temporal structure. We also propose using \texttt{TexStat} as a validation metric alongside Frechet Audio Distances (FAD) to evaluate texture sound synthesis models. In addition to \texttt{TexStat}, we present \texttt{TexEnv}, an efficient, lightweight and differentiable texture sound synthesizer that generates audio by imposing amplitude envelopes on filtered noise. We further integrate these components into \texttt{TexDSP}, a DDSP-inspired generative model tailored for texture sounds. Through extensive experiments across various texture sound types, we demonstrate that \texttt{TexStat} is perceptually meaningful, time-invariant, and robust to noise, features that make it effective both as a loss function for generative tasks and as a validation metric. All tools and code are provided as open-source contributions and our PyTorch implementations are efficient, differentiable, and highly configurable, enabling its use in both generative tasks and as a perceptually grounded evaluation metric. 
\end{abstract}

\section{Introduction}\label{sec:introduction}
Defining audio textures is a complex problem that has been explored by various authors. The concept originated as an analogy to visual textures. In \cite{julesz1962}, Julesz, one of the pioneers in this field, proposed the so-called "Julesz conjecture," suggesting that humans cannot distinguish between visual textures with similar second-order statistics. This hypothesis was later disproven in \cite{caelli1978}, but whose statistical perspective remains influential in texture analysis (see \cite{humeau2019}) and synthesis (see \cite{mcdermott2009}, \cite{mcdermott2011}, and \cite{mcdermott2013}). This perspective is also foundational for this work. 

Regarding the auditory domain, an early definition of audio textures was introduced in \cite{saintarnaud1995}. The authors described them as patterns of basic sound elements, called atoms, occurring in structured but potentially random arrangements. These high-level patterns must remain stable over time while being perceivable within a short time span. Rosenthal et al. \cite{rosenthal1998} later expanded this idea, defining audio textures as a two-level structure: atoms forming the core elements and probability-based transitions governing their organization. Recent research has refined these ideas, but the balance between short-term unpredictability and long-term stability remains a common thread. In this matter, Wyse et al. \cite{wyse2022} points out that sound texture's complexity and unpredictability at one level is combined with the sense of eternal sameness at another.

Many approaches have been taken regarding the synthesis and re-synthesis of texture sounds. These approaches can be classified in various ways. For example, Sharma et al. \cite{sharma2022} separates them into three categories: granular synthesis, which corresponds to the concatenation of already existing sounds, as in the pioneering works \cite{saintarnaud1995_synthesis}, \cite{Dubnov2002},
and \cite{Schwarz2004}; model-based synthesis, which relies on either improved time-frequency representation models, as in \cite{DiScipio2002}, physical models, as in \cite{OBrien2002}, and/or physiological/statistical models, as in the foundational series of articles \cite{mcdermott2009}, \cite{mcdermott2011}, and \cite{mcdermott2013}; and finally, deep learning-based models, encompassing various contemporary machine learning techniques, as in \cite{Caracalla2019}, 
\cite{wyse2022}, and \cite{NoiseBandNets}.

The sound resynthesis task involves recreating a given sound through analysis and synthesis. While sound reconstruction can be viewed as an exact replication of the original sound, in the case of texture-based resynthesis, it is often sufficient to generate a sound that is perceptually similar to the original, and thus, the resulting sound may differ significantly in its detailed time structure from the original one. Moreover, Caracalla et al. \cite{Caracalla2019} state that typically the texture sound resynthesis goal is to "create sound that is different from the original while still being recognizable as the same kind of texture." In the context of deep learning, this goal can be approached in various ways. For instance, models with inherent stochasticity naturally generate a "similar enough" sound, drawing from the set of sounds they can produce, which is influenced by their biases. Conversely, even in the absence of stochasticity, if the model's loss function is perceptually grounded, it will inevitably focus solely on that instead of exact replication.

An example of a model that performs resynthesis by accurately matching the spectrum of a texture sound is NoiseBandNet \cite{NoiseBandNets}, a Differentiable Digital Signal Processing (DDSP)-based architecture that indirectly follows this approach by using a multi-scale spectrogram loss applied to small time scales (starting at approximately $0.7$ ms) and small sound windows (around $180$ ms). On the side of perceptual/features-based resynthesis using deep learning, \cite{Caracalla2019} employed a loss function that compares the Gram matrices of two sets of features computed from pretrained Convolutional Neural Networks (CNNs). Another example comes from \cite{wyse2022} and its use of Generative Adversarial Networks (GANs), where the objective of the loss function is to train the model to generate sounds that can deceive a classifier, thereby biasing the training in a more nuanced manner than direct signal comparison.

In the context of texture sound analysis and resynthesis, this paper has three main goals. First, we introduce \texttt{TexStat}, a loss function grounded in human perception of texture sounds. This function builds upon the pioneering work of McDermott and Simoncelli \cite{mcdermott2009}, \cite{mcdermott2011}, \cite{mcdermott2013}, and is made available as an efficient, open-source implementation in PyTorch. Second, we present \texttt{TexEnv}, a simple and efficient noise-based texture sound synthesizer developed in synergy with \texttt{TexStat}. Finally, to evaluate both \texttt{TexStat} and \texttt{TexEnv}, we introduce \texttt{TexDSP}, a DDSP-based architecture trained across multiple scenarios, demonstrating the effectiveness of our proposed components.

In Section \ref{sec:loss}, we formally define the loss function \texttt{TexStat} together with its potential variations. In Section \ref{sec:synth}, we briefly introduce the \texttt{TexEnv} synthesizer, highlighting both its capabilities and its limitations. In Section \ref{sec:arch}, we present the \texttt{TexDSP} architecture as a showcase for the tools introduced earlier. In Section \ref{sec:results}, a series of experiments demonstrating the tools introduced in this paper are presented. Finally, in Section \ref{sec:conclusions}, we summarize the tools and results discussed in this work and outline some future directions for research.

\section{\texttt{TexStat}: A Loss Function Specifically Tailored for Texture Sounds}\label{sec:loss}
In this section, we discuss some desirable properties of a texture sound loss function and then introduce \texttt{TexStat}, a loss function specifically designed for texture sounds that fulfills, to some extent, the properties outlined here.

\subsection{What Should a Texture Loss Be?}\label{subsec:good_loss}
In the context of deep learning, solving a task—whatever the task may be—is framed as an optimization problem. Such an optimization problem typically involves minimizing a loss function, with the expectation that doing so will lead to solving the proposed task. In this scenario, the loss function must be able to represent, to some extent, the task to be solved.

The texture sound generation task can be described as a process in which a sound is created so that it is recognized as belonging to a particular type of texture. However, it does not need to be—and in fact, it is preferable that it is not—identical to the input or dataset examples. Additionally, depending on the context, it may also be desirable to generate non-repetitive sound over time, with the sound generation process allowing for some degree of temporal control.

This task description, together with our current understanding of texture sounds, leads us to the following desirable properties for a loss function specifically tailored for texture sounds.

\sloppy\noindent\textbf{Overall Structure Focus:} Texture sounds resemble filtered noise over short periods but are stable and recognizable over time. Thus, a loss function should prioritize long-term characteristics, like rhythm, pitch variations, or granular shifts, rather than specific and detailed temporal structure.

\noindent\textbf{Stochastic Focus:} Following \cite{julesz1962} and \cite{rosenthal1998}, texture sounds arise from stochastic processes in time and timbre. A suitable loss function should capture these statistical properties to recognize similarity between sounds corresponding to the same type of texture.

\noindent\textbf{Perceptual Foundation:} The loss function must capture perceptual similarity, ensuring that two sounds considered perceptually similar are treated as such. This involves leveraging psychoacoustic principles to align with human auditory perception.

\noindent\textbf{Time Invariance:} Texture sounds can be chopped or rearranged with minimal perceptual impact. The loss function should thus exhibit time invariance, allowing rearrangement without significantly altering the sound’s characteristics.

\noindent\textbf{Noise Stability:} Texture sounds can tolerate noise, so the loss function should be robust to subtle noise variations, preserving core texture features even with low to mid level disturbances.

\noindent\textbf{Flexibility:} A highly restrictive loss function risks overfitting, leading to audio too similar to the training data. It should encourage creative variation, generating novel yet perceptually consistent sounds that retain defining textural characteristics.

\subsection{\texttt{TexStat} Formal Definition}
According to Julesz's conjecture and McDermott and Simoncelli's synthesis approach, the nature of textures can be understood through direct comparison of their statistics. McDermott and Simoncelli's synthesis approach \cite{mcdermott2009}, \cite{mcdermott2011}, and \cite{mcdermott2013} involves the imposition of a set of statistics precomputed for a given texture sound using numerical methods, suggesting the possibility of using that exact set of statistics as a feature vector for comparison.

In this work, we introduce \texttt{TexStat}, a loss function that operates by directly comparing a slight variation of a superset of the summary statistics used by McDermott and Simoncelli.

In the following, we formally introduce the necessary tools, preprocessing steps, and sets of summary statistics related to the computation of \texttt{TexStat}. For further details on summary statistics' perceptual importance, we refer to \cite{mcdermott2011} and \cite{mcdermott2013}.

\noindent\textbf{Preliminary Tools:} At the core of \texttt{TexStat} there are a series of subband decompositions. Following the original work of McDermott and Simoncelli, we consider the following: a Cochlear Filterbank, which is an Equivalent Rectangular Bandwidth (ERB) filterbank \cite{erb1}, \cite{erb2} $F=\{f_j\}_{j=1}^{N_F}$ made up of $N_F$ filters; and a Modulation Filterbank, which is a Logarithmic Filterbank $G=\{g_k\}_{k=1}^{N_G}$ made up of $N_G$ filters.

\noindent\textbf{Preprocessing:} Given a signal $s$, the Cochlear filterbank $F$ is used to compute a subband decomposition of it by $s_j= s * f_j,$ $j=1,\dots,N_F$. Then, the amplitude envelope of each one of these subbands is computed using the Hilbert Transform
$e_j=|s_j+ iH(s_j)|,$ $j=1,\dots,N_F$. Finally, the Modulation Filterbank $G$ is used to compute a subband decomposition of each one of the envelopes $e_j$, that is, $m_{j,k}=g_k(e_j), j=1,\dots,N_F,k=1,\dots,N_G$.

\noindent\textbf{Statistics sets:} The first set of statistics is comprised of the first $L$ normalized moments of the signals $e_j$, that is,
$S_{1,l,j} = M_l(e_j)$ for $j=1,\dots,N_F$ and $l=1,\dots,L$, where $M_1(X)=\text{E}(X)=\mu$, $M_2(X)=V(X)/\mu^2=\sigma^2/\mu^2$ and 
$$M_l(X)=\frac{\text{E}(X-\mu)^l}{\sigma^l},\quad l=3,\dots, L.$$
Note that the vectors $S_{1,l}\in \mathbb{R}^{N_F}$ correspond to different moments and hence they are prone to have values that vary in orders of magnitude. In order to fix this issue, the first set of statistics correspond to a weighted concatenation of the vectors corresponding to different moments
$$S_1(s) = \text{concat}(\alpha_1S_{1,1}, \dots, \alpha_N S_{1,L}) \in\mathbb{R}^{L\cdot N_F}.$$

The second set of statistics corresponds to the Pearson correlation between different amplitude envelopes $e_j$, that is, 
$$     S_2(s) = \text{vech}(\text{corr}(e_1,\dots,e_{N_F})) \in\mathbb{R}^{T_{N_F-1}}, $$
where $\text{vech}(\cdot)$ corresponds to the half-vectorization operator, $\text{corr}(\cdot)$ to the correlation matrix and $T_n$ to the $n$-th triangular number.

The third set of statistics corresponds roughly to the proportion of energy in each modulation band, that is,
\begin{equation*}
\begin{array}{rrll}
S_{3,j} &=& \displaystyle\frac{\left(\text{V}(m_{j,1})^{1/2},\dots,\text{V}(m_{j,N_G})^{1/2}\right)}{\text{V}(e_j)^{1/2}}&\in\mathbb{R}^{N_G}\\
S_{3}(s)&=&\text{concat}(S_{3,1},\dots,S_{3,N_F})&\in\mathbb{R}^{N_G\cdot N_F}.
\end{array}
\end{equation*}

The fourth set of statistics corresponds to the Pearson correlations between modulation subbands corresponding to the same amplitude envelope, that is,
\begin{equation*}
\begin{array}{rrcl}
S_{4,j} &=& \text{vech}(\text{corr}(m_{j,1},\dots,m_{j,N_G}))&\in\mathbb{R}^{T_{N_G-1}}\\
S_4(s) &=& \text{concat}(S_{4,1},\dots,S_{4,N_F})&\in\mathbb{R}^{N_F\cdot T_{N_G-1}}.
\end{array}
\end{equation*}

Finally, the fifth set of statistics corresponds to the Pearson correlation between modulation subbands corresponding to the same band but different amplitude envelopes, that is,
\begin{equation*}
\begin{array}{rrcl}
S_{5,k} &=& \text{vech}(\text{corr}(m_{1,k},\dots,m_{N_F,k}))&\in\mathbb{R}^{T_{N_F-1}}\\
S_5(s) &=& \text{concat}(S_{5,1},\dots,S_{5,N_G})&\in\mathbb{R}^{N_G\cdot T_{N_F-1}}.
\end{array}
\end{equation*}

Now that we have the five sets of statistics well-defined, we can finally define the \texttt{TexStat} loss function.\\

\noindent\textbf{Definition 1. (\texttt{TexStat} loss function)} The \texttt{TexStat} loss function is defined as
$$ \mathcal{L}_{\alpha,\beta}(x,y) = \sum_{j=1}^5 \beta_j \cdot \text{MSE}(S_j(x),S_j(y)),$$
where $\alpha\in\mathbb{R}^L$ and $\beta\in\mathbb{R}^5$ are parameters, $x,y$ are a pair of signals, and $S_1(x),S_2(y),\dots,S_5(x),S_5(y)$ are the summary statistics vectors defined above.\\

Before continuing, it is important to mention that although the \texttt{TexStat} loss function is strongly based on the work of McDermott and Simoncelli, several changes and additions were made to make it more suitable for machine learning tasks. Most of these changes involve adaptations that ensure control over the number of statistics and their weights during training. The number of statistics is important, as it would be counterproductive to compute more statistics than the size of the window used. Furthermore, weighting the statistics provides a way to balance their contribution to the loss. The main changes and their impact are outlined in Table \ref{tab:mcdermott_vs_texstat}.
\begin{table}[h]
    \centering
    \small
    \renewcommand{\arraystretch}{1.2} 
    \setlength{\tabcolsep}{2pt} 
    \begin{tabularx}{\columnwidth}{
        >{\raggedright\arraybackslash}p{0.22\columnwidth} 
        >{\raggedright\arraybackslash}p{0.33\columnwidth} 
        >{\raggedright\arraybackslash}X}
        \toprule
        \textbf{} & \textbf{McDermott and Simoncelli's Summary Statistics} & \textbf{\texttt{TexStat} Summary Statistics} \\
        \midrule
        Filterbanks & Fixed type, frequency range, and size of filterbanks for their experiments. & Variable type, frequency range, and size of filterbanks for controllability. \\
        Statistical Moments & Fixed number of statistical moments. & Variable number of statistical moments. Useful for compensating small filterbanks. \\
        Modulation Band Correlations & Only some correlations are computed. & Multiple correlations were avoided in the original work for computational efficiency; however, with modern GPU computations, this doesn't make a significant difference. \\
        Compressive Non-linearity & Applies one to the amplitude envelopes following past auditory models. & Removes compressive non-linearity for gradient stability. \\
        Weights & Summary statistics are not weighted. & Variable weights to control the importance of certain sets of statistics during training and to avoid overflow (especially in $S_1$). \\
        \bottomrule
    \end{tabularx}
    \caption{Summary statistics comparison between McDermott and Simoncelli's work, and those used by \texttt{TexStat}.}
    \label{tab:mcdermott_vs_texstat}
\end{table}

\subsection{\texttt{TexStat} Properties}
As discussed in Subsection \ref{subsec:good_loss}, there are several desirable properties that a texture sound loss function should have, and we argue that, when used correctly, \texttt{TexStat} addresses them all.\\
First, \texttt{TexStat} can be utilized on arbitrarily large frames of audio, and in any case, all summary statistics are directly influenced by the entire signal, which we argue implies a focus on the overall structure. Moreover, since it is built on a statistical and perceptual model, we also argue that its focus is on the stochastic properties of the signal, and that it has a strong foundation in perception.\\
Overall, one could argue that the operations involved in the computation of summary statistics are quite stable with respect to the addition of low-amplitude white noise. This will be empirically demonstrated in Subsection \ref{subsec:experiments_properties}. Moreover, if $s \in C(\mathbb{R})$ is a continuous infinite signal whose summary statistics exist (for example, if it belongs to the Schwarz class $s \in \mathcal{S}(\mathbb{R})$), there are no operations in the process of computing $S_1(s), \dots, S_5(s)$ that can be affected by time shifting, i.e., $\hat{s}(t) = s(t - t_0)$. This is, of course, not the case for discrete finite signals, where time shifting, $\hat{s}[t] = s[t - t_0 \mod \text{len}(s)]$, might introduce a click sound at the beginning, which would affect the spectrum and, consequently, the subband decomposition. However, this is not a significant issue for noisy signals, as adding a click will not strongly impact their spectrum. This will also be further explored in Subsection \ref{subsec:experiments_properties}.

Finally, given a choice of parameters that does not generate an excessive number of summary statistics in relation to the samples used in the original signal, there are generally many signals that, when compared using \texttt{TexStat}, will result in low loss values. For example, all chopped and reordered versions of the same signal will typically yield similar summary statistics, and hence would be close in the sense of the \texttt{TexStat} distance. We claim that this suggests flexibility for the \texttt{TexStat} loss function.

\subsection{Capabilities and Limitations}\label{subsec:cap_lim}
The \texttt{TexStat} loss function is based on the direct comparison of summary statistics, meaning that two sounds whose summary statistics are similar will not be recognized as different by this loss function. In their original work, McDermott and Simoncelli imposed the summary statistics of a series of sounds onto white noise and found that, although they successfully synthesized a range of sounds with good results, this process was unable to generate convincing sounds for certain types of timbres. This was because the summary statistics were not sufficient to fully characterize those types of sounds. Regarding the sounds that couldn't be fully captured by the summary statistics, McDermott and Simoncelli stated that "they fall into three general classes: those involving pitch (e.g., railroad crossing, wind chimes, music, speech, bells), rhythm (e.g., tapping, music, drumming), and reverberation (e.g., drumbeats, firecrackers)." These capabilities and limitations are naturally inherited by the \texttt{TexStat} loss function, and both effective and ineffective examples are shown in Subsection \ref{subsec:experiments_texdsp} to make transparent this loss' limitations.

\subsection{Usage as a Loss and/or Evaluation Metric}\label{subsec:loss_eval_usage}
The \texttt{TexStat} loss function introduced here is differentiable, making it suitable for use as a loss function in machine learning applications. Moreover, together with this article we release an efficient and open-source PyTorch implementation.

The number of computations required to run both \texttt{TexStat} and its gradients are relatively large compared to other loss functions, such as the Multi-Scale Spectral (MSS) Loss function. However, this can be mitigated by adjusting the frame size, as summary statistics are intended to be used on large audio frames. For example, most models evaluated in Section \ref{subsec:experiments_texdsp} used a frame size corresponding to approximately $1.5$ seconds of audio at a framerate of $44100$ Hz. In contrast, architectures like NoiseBandNets \cite{NoiseBandNets} use the MSS on frames of up to $0.18$ seconds. Additionally, the increased computational cost also comes with higher memory usage, especially when increasing $N_F$ and $N_G$, which can significantly impact memory requirements. 

Given the timbre limitations of the summary statistics discussed earlier, we believe that, in order to fully guide the learning process of a general generative model, additional losses should be introduced to exert full control over different types of sounds, such as pitched or rhythmic sounds. In such cases, \texttt{TexStat} can be regarded as a regularizer that aids in guiding the timbre learning of texture sounds.

In some instances, \texttt{TexStat} may not be suitable as a loss function for the reasons mentioned earlier. In these cases, one can use other losses to guide the learning process and, if appropriate for the task, employ either \texttt{TexStat} as an evaluation metric. To facilitate this, we propose a fixed set of parameters, including filterbank types, filterbank sizes, and values for $\alpha$ and $\beta$, which have been proven useful for texture sound synthesis in the past and that were tested in this article's experiments. These parameters ensure comparability, and we also provide a list of precomputed values for interpretability. All of this can be found in the repository for the \texttt{TexStat} loss function. Moreover, since the \texttt{TexStat} loss function is essentially a direct comparison of summary statistics, one could view this set of statistics as a fully interpretable feature vector. This feature vector can thus be used as the basis for other evaluation metrics, such as Frechet Audio Distance (FAD) \cite{FAD}.

Our repository comes with a simple to use implementation of this method that uses a subset of the summary statistics here proposed and extensive experimentation to prove this concept can be found in Subsections \ref{subsec:experiments_evaluation} and \ref{subsec:experiments_texdsp}.

\section{\texttt{TexEnv}: A Differentiable Signal Processor Tailored for Texture Sounds}\label{sec:synth}
The foundations of \texttt{TexStat} are based on the idea that summary statistics of amplitude envelopes derived from a low-size filterbank subband decomposition are sufficient to compare certain types of texture sounds. Implicit in this concept is the fact that directly imposing amplitude envelopes on a subband decomposition of white noise is one method of resynthesizing texture sounds while preserving their summary statistics. In the context of this work, two questions arise: How can we efficiently and differentiably create amplitude envelopes from a sequence of parameters? How can we efficiently and differentiably impose amplitude envelopes?

Creating amplitude envelopes from scratch can be done in multiple ways. For this synthesizer, we chose to use the Inverse Fast Fourier Transform (IFFT) because it is differentiable and can be computed efficiently to generate cyclic functions. For the imposition process, we fixed precomputed white noise, decomposed it using a filterbank, and then normalized the amplitude envelope of each subband. This procedure generates an object we call the \textit{seed}, which serves as a "source of deterministic randomness" and can be used to impose amplitude envelopes via simple multiplication. Algorithm \ref{alg:TexEnv} outlines this synthesis method.

\begin{algorithm}[h]
	\caption{The \texttt{TexEnv Synth}} 
        \label{alg:TexEnv}
        \textbf{Input:} Let $F$ be a filterbank of size $N_F$, $(s_1, \dots, s_{N_F})$ a \textit{seed} generated from $F$, $p_1,\dots,p_{N_F}\in\mathbb{C}^{N_P}$ a set of complex vector parameters and $N$ the length of the signal to be generated.\\
        \textbf{Output:} A texture sound synthesized $y\in\mathbb{R}^N$.
	\begin{algorithmic}[1]
            \State For each $j=1,\dots,N_F$ construct a spectrum as
            $$
            A_j = \text{concat}( p_{j,0}, \dots, p_{j,N_P-1}), \mathbf{0}, (\overline{p}_{j,N_P-1}, \dots, \overline{p}_{j,1})),
            $$
            where $\mathbf{0}$ is a null vector of size $N-2N_P+1$.
            \State For each $j=1,\dots,N_F$ construct a real signal using the Inverse Discrete Fourier Transform (IDFT)
            $$ a_j = \text{IDFT}(A_j).$$
            \State Impose the signals $a_1,\dots,a_{N_F}$ as amplitude envelopes on the seed and sum up to generate the final signal
            $$ y = \sum_{j=1}^{N_F} s_j \odot  a_j.$$
            \State \Return The synthesized signal $y$.
	\end{algorithmic}
\end{algorithm}

\section{\texttt{TexDSP}: A DDSP-Based Architecture Tailored for Texture Sounds}\label{sec:arch}
In this section we introduce \texttt{TexDSP}, a relatively simple texture sound generative model based on DDSP \cite{engel2020} that showcases the capabilities of \texttt{TexStat} and \texttt{TexEnv}.

\begin{figure*}[ht]
    \centering
    \includegraphics[width=0.9\textwidth]{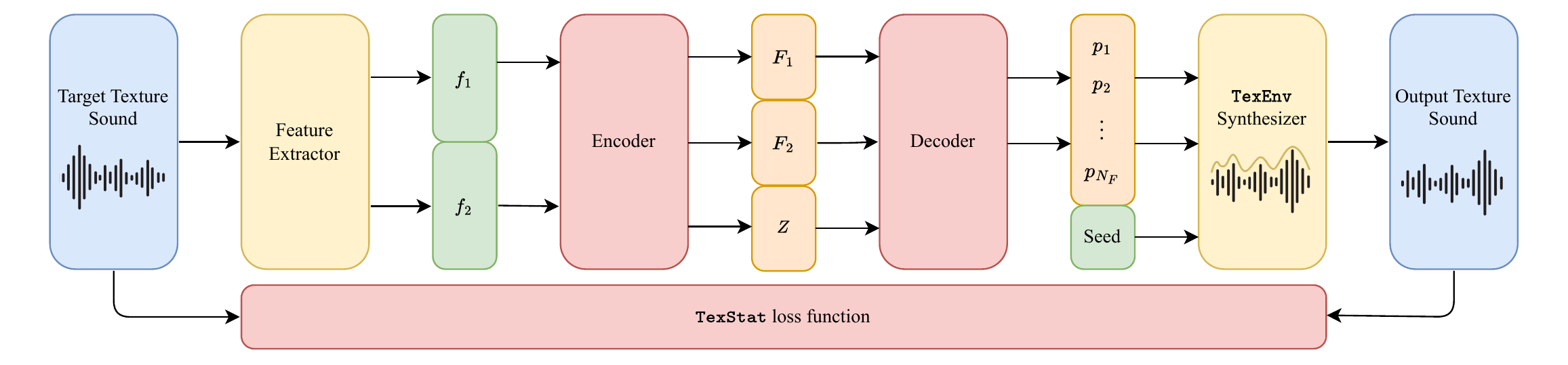}
    \caption{\texttt{TexDSP} architecture. Prechosen features are computed and are used to run the model. The encoder adds complexity and entangles this features into the latent representation $L=(F_1,F_2,Z)$. The decoder transforms this representation into a set of complex parameters that are used to run the \texttt{TexEnv} synthesizer. Finally, the output signal is compared to the original one using the \texttt{TexStat} loss function.}
    \label{fig:arch}
\end{figure*}

The original DDSP model requires an encoder, decoder, signal processor, and loss function, with each component designed to generate small frames of pitched sounds based solely on pitch and loudness features. However, in the context of texture sound generation, the task is quite different, necessitating several modifications. The changes we propose ensure that the model's objective is to learn high-level statistical patterns, rather than aiming for perfect (frequency domain) reconstruction, as is the case with the original DDSP architecture. The final architecture is shown in Figure \ref{fig:arch}, and the changes made are briefly outlined here.

\subsection{Encoder and Decoder}
As in the original DDSP architecture, the encoding process involves both feature extraction and a more complex transformation of these features. In this case, we used different pairs of features that can be more informative for texture sounds than pitch and loudness. These features include spectral mean (frequency centroid), spectral standard deviation, energy in each band of a subband decomposition, and onset rate. To increase the complexity of the feature extraction process, we followed the original approach: first, we applied a Multi-Layer Perceptron (MLP) to each feature $F_j = \text{MLP}(f_j \mid \varepsilon_j^{\text{enc}})$, and then concatenated the results with the output of a Gated Recurrent Unit (GRU) applied to the same features $Z=\text{GRU}(F_1, F_2 \mid \varphi)$, yielding the latent representation $L = \text{concat}(F_1, F_2, Z)$. The decoding process involves transforming the latent representation $L$ through another MLP and an additional layer to obtain a polar representation of the parameters used by the signal processor, i.e., 
\begin{equation*}
\begin{aligned}
\rho &= \sigma(A \cdot \text{MLP}(L \mid \varepsilon^{\text{dec}})) \\
\theta &= 2\pi \sigma(B \cdot \text{MLP}(L \mid \varepsilon^{\text{dec}}))
\end{aligned}
\end{equation*}
where $A,B$ are real-valued matrices to be learned, and $\sigma(\cdot)$ denotes the sigmoid function applied component-wise.

\subsection{Signal Processor and Loss Function}
The original DDSP model used a Spectral Modeling Synthesis (SMS) \cite{serra1990} synthesizer, which is well-suited for pitched sounds, and a Multi-Scale Spectrogram (MSS) Loss, which is effective for perfect reconstructions. Since the goal of this architecture is to test the capabilities of \texttt{TexStat}, we opted for our signal processor, as it was designed to work synergistically with our loss function. Both \texttt{TexEnv} and \texttt{TexStat} are built around a filterbank, and to optimize synergy, this filterbank is shared between the two.

\section{Experiments and Results}\label{sec:results}
In this section, we briefly explain a series of experiments conducted to provide proof of concept for the models proposed in this work. For these experiments, we hand-curated  \texttt{MicroTex}\footnote{\texttt{MicroTex} HuggingFace repository: \href{https://huggingface.co/datasets/cordutie/MicroTex}{cordutie/MicroTex}.}, a dataset made from a selection of texture sounds from the following sources: the BOReilly dataset, containing textures made using analog synthesizers; Freesound \cite{freesound}, which contains environmental sounds; and synthetically generated data using Syntex \cite{syntex}. All experiments can be found in their respective repositories and are fully replicable\footnote{Find access to all repositories, experiments and sound examples in this article's webpage: \href{https://cordutie.github.io/ddsp_textures/}{cordutie.github.io/ddsp\_textures/}.}. 

\subsection{\texttt{TexStat} Properties Tests}\label{subsec:experiments_properties}
Two desirable properties proposed for a loss function tailored to texture sounds are stability under time shifting and robustness to added noise. To test these properties in the \texttt{TexStat} loss function, we computed the loss between the sounds in the Freesound class of \texttt{MicroTex} and their corresponding transformations using various parameters. We focused on the Freesound class of \texttt{MicroTex}, as it contains the most representative examples of environmental texture sounds and includes recordings that are long and dynamic enough to allow meaningful time shifting and noise addition. The other two classes were excluded, as their sounds are either too short or too silent, making such transformations impractical without introducing substantial alterations. The experiment was also run with the MSS loss for comparison and some of the results can be found in Table \ref{tab:exp_prop_time_noise}.
\begin{table}[h]
    \centering
    \small
    \renewcommand{\arraystretch}{1.2}
    \setlength{\tabcolsep}{2pt}
    \begin{tabularx}{\columnwidth}{
        >{\raggedright\arraybackslash}p{0.22\columnwidth}
        >{\centering\arraybackslash}X
        >{\centering\arraybackslash}X 
        >{\centering\arraybackslash}X
        >{\centering\arraybackslash}X
        >{\centering\arraybackslash}X
        >{\centering\arraybackslash}X
    }
    \toprule
    & \multicolumn{3}{c}{\texttt{TexStat}} & \multicolumn{3}{c}{MSS} \\
    \cmidrule(r{2pt}){2-4} \cmidrule(l{2pt}){5-7}
    Transformation & $10\%$ & $30\%$ & $50\%$ & $10\%$ & $30\%$ & $50\%$ \\
    \midrule
    \noalign{\vskip 2pt} 
    Time-Shift &
    \shortstack{$0.04$ \;\;\\ $\pm~0.03$} &
    \shortstack{$0.04$ \;\;\\ $\pm~0.03$} &
    \shortstack{$0.04$ \;\;\\ $\pm~0.03$} &
    \shortstack{$6.09$ \;\;\\ $\pm~1.22$} &
    \shortstack{$6.27$ \;\;\\ $\pm~1.38$} &
    \shortstack{$6.29$ \;\;\\ $\pm~1.41$} \\
    \noalign{\vskip 4pt} 
    Noise-Add &
    \shortstack{$2.08$ \;\;\\ $\pm~1.99$} &
    \shortstack{$2.51$ \;\;\\ $\pm~2.21$} &
    \shortstack{$2.65$ \;\;\\ $\pm~2.27$} &
    \shortstack{$11.79$ \;\;\\ $\pm~4.91$} &
    \shortstack{$16.84$ \;\;\\ $\pm~5.92$} &
    \shortstack{$19.57$ \;\;\\ $\pm~6.26$} \\
    \noalign{\vskip 2pt} 
    \bottomrule
    \end{tabularx}
    \caption{Loss measurements ($\mu\pm\sigma$) between sounds in the Freesound class of \texttt{MicroTex} and their corresponding time-shifted and noise-added transformations. Time shift is expressed as a percentage of the total signal duration, and noise percentage are defined by their maximum amplitude relative to the original signal. All measurements were computed over one-second segments for each of the sounds mentioned above. For reference, all satisfactory models trained using \texttt{TexStat} converged to loss values below $3$, whereas evaluations using MSS typically yield reasonable values below $10$.}
    \label{tab:exp_prop_time_noise}
\end{table}

The results demonstrate that \texttt{TexStat} exhibits strong stability under both time shifting and noise addition, incurring a consistent penalty for time shifts and a sublinear increase in penalty as noise levels increases.

\subsection{\texttt{TexStat} Benchmarks}\label{subsec:experiments_benchmarks}
To benchmark the computational requirements of \texttt{TexStat}, we evaluated its computation time, gradient descent time, and GPU memory usage. These measurements were conducted multiple times, recording the time taken for loss computation and optimization while tracking memory allocation. The results are presented in Table \ref{tab:exp_computation}, along with the values for other typical losses.
\begin{table}[h]
    \centering
    \small
    \renewcommand{\arraystretch}{1.2}
    \setlength{\tabcolsep}{2pt}
    \begin{tabularx}{\columnwidth}{
    >{\raggedright\arraybackslash}p{0.2\columnwidth} 
    >{\raggedright\arraybackslash}X
    >{\raggedright\arraybackslash}X 
    >{\raggedright\arraybackslash}X}
        \toprule
        Loss & Forward pass time (ms) & Backward pass time (ms) & Memory usage (mb)\\
        \midrule
        \texttt{TexStat} & $93.5 \pm 0.4$ & $154.6 \pm 0.4$ & $0.84 \pm 2.5$\\
        MSS              & $3.9 \pm 0.3$  & $8.5 \pm 0.3$   & $0.85 \pm 2.6$\\
        MSE              & $0.2 \pm 0.3$  & $0.2 \pm 0.1$   & $1.7 \pm 5.0$\\
        MAE              & $0.1 \pm 0.0$  & $0.2 \pm 0.1$   & $0.8 \pm 2.5$\\
        \bottomrule
    \end{tabularx}
    \caption{Measurements regarding computation time, gradient computation time, and memory usage ($\mu\pm\sigma$) in batches of $32$ signals of size $65536$ (around $1.5$s at a sample rate of $44100$\,Hz). The losses studied were \texttt{TexStat}, Multi-Scale Spectrogram (MSS), Mean Squared Error (MSE), and Mean Absolute Error (MAE). All measurements were done using CUDA on an RTX 4090 GPU.}
    \label{tab:exp_computation}
\end{table}

The results show that, as expected, the \texttt{TexStat} loss function is slower than other less specific losses, but it uses a similar amount of memory.

\subsection{\texttt{TexStat} as an Evaluation Metric}\label{subsec:experiments_evaluation}
In order to test \texttt{TexStat} summary statistics as a powerful representation that can be used in metrics like FAD, we conducted the following experiment. First, all data in the three selections of the \texttt{MicroTex} dataset were segmentated and both their summary statistics and VGGish \cite{vggish} embeddings were computed. Then, a downstream classifier (MLP with hidden layers 128, 64) was trained in both cases. A summary of the results can be found in Table \ref{tab:exp_eval}.
\begin{table}[h]
    \centering
    \small
    \renewcommand{\arraystretch}{1.2}
    \setlength{\tabcolsep}{2pt}
    \begin{tabularx}{\columnwidth}{
    >{\raggedright\arraybackslash}p{0.16\columnwidth}
    >{\raggedright\arraybackslash}p{0.16\columnwidth}
    >{\raggedright\arraybackslash}X 
    >{\raggedright\arraybackslash}X
    >{\raggedright\arraybackslash}X 
    >{\raggedright\arraybackslash}X
    }
        \toprule
        Model            & Selection & Accuracy & Precision & Recall & F1\\
        \midrule
        \texttt{TexStat} & BOReilly  & $\mathbf{0.94}$  & $\mathbf{0.94}$  & $\mathbf{0.94}$  & $\mathbf{0.94}$\\
        VGGish           & BOReilly  & $0.71$           & $0.73$           & $0.71$           & $0.71$\\
        \texttt{TexStat} & Freesound & $\mathbf{0.99}$  & $\mathbf{0.99}$  & $\mathbf{0.99}$  & $\mathbf{0.99}$\\
        VGGish           & Freesound & $0.98$           & $0.99$           & $0.98$           & $0.98$\\
        \texttt{TexStat} & Syntex    & $\mathbf{1.0}$   & $\mathbf{1.0}$   & $\mathbf{1.0}$   & $\mathbf{1.0}$\\
        VGGish           & Syntex    & $0.95$           & $0.95$           & $0.95$           & $0.94$\\
        \bottomrule
    \end{tabularx}
    \caption{Classification performance of equivalent models trained on both our proposed feature vector and VGGish embeddings.}
    \label{tab:exp_eval}
\end{table}

The results indicate that in the context of texture sounds, summary statistics are strictly more informative than general-purpose embeddings such as VGGish.

\subsection{Texture resynthesis using \texttt{TexEnv}}\label{subsec:experiments_resynthesis}
Extensive exploration using the \texttt{TexEnv} synthesizer in resynthesis tasks employing a signal processing-based parameter extractor was conducted to better understand its limitations and overall behavior. A summary of sound examples can be found on this article's webpage. Some of our key findings were as follows: water-like sounds such as flowing water, rain, and continuous bubbling do not benefit from larger parameter sets, but do benefit from larger filterbanks. In contrast, crackling sounds like fireworks or bonfires benefit from larger parameter sets, but not as much from larger filterbanks. These insights were crucial in determining the optimal parameters for the models trained later.

\subsection{\texttt{TexDSP} Trained Models}\label{subsec:experiments_texdsp}
To showcase the capabilities of \texttt{TexStat}, we trained a set of \texttt{TexDSP} models using different parameters, with \texttt{TexStat} as the sole loss function to guide the learning process. The details and results for some of these models are presented below.

\noindent\textbf{Training Details:} A curated set of sounds representing different classes of texture sounds from Freesound was used for each model. Each model employed different parameters tailored to the specific texture type. These parameters were chosen based on the resynthesis exploration discussed in Subsection \ref{subsec:experiments_resynthesis}. The number of layers in the MLPs for both the encoder and decoder was limited to a maximum of $3$, with the number of parameters capped at $512$. This configuration ensured that the resulting models, even when combined with the seed used for the filterbank, remained under $25$ MB and could, if necessary, be ported to a real-time environment. The \texttt{TexStat} $\alpha$ and $\beta$ parameters were set to the default values proposed in our repository, and all models used the same optimizer, training for up to $1500$ epochs with early stopping enabled. Additionally, for each \texttt{TexDSP} model trained, a corresponding NoiseBandNet model was also trained using default parameters for comparison.


\noindent\textbf{Validation Method:} To evaluate model performance, we resynthesized a subset of the dataset, excluded from training, for both the \texttt{TexStat} and NoiseBandNet models. We then segmented the original and resynthesized signals, and measured Fréchet Audio Distance (FAD) using both VGGish embeddings and our custom summary statistics, along with frame-level \texttt{TexStat} and MSS losses. For the latter, we report the mean and standard deviation across all segments. Results are presented in Table~\ref{tab:exp_validation}.

\begin{table*}[h]
    \centering
    \small
    \renewcommand{\arraystretch}{1.2}
    \setlength{\tabcolsep}{2pt}
    \begin{tabularx}{\textwidth}{
    >{\arraybackslash}p{0.15\columnwidth}
    >{\centering\arraybackslash}X
    >{\centering\arraybackslash}X
    >{\centering\arraybackslash}X
    >{\centering\arraybackslash}X
    >{\centering\arraybackslash}X
    >{\centering\arraybackslash}X
    >{\centering\arraybackslash}X
    >{\centering\arraybackslash}X
    }
    \toprule
    \multicolumn{1}{c}{} & 
    \multicolumn{4}{c}{\textbf{FAD}} & 
    \multicolumn{4}{c}{\textbf{Loss metrics}} \\
    \cmidrule(r{4pt}){2-5} \cmidrule(l{4pt}){6-9}
    Texture Sound & VGGish \texttt{TexDSP} & VGGish NoiseBandNet & Ours\; \texttt{TexDSP} & Ours NoiseBandNet & \texttt{TexStat} \texttt{TexDSP} & \texttt{TexStat} NoiseBandNet & MSS \;\texttt{TexDSP} & MSS NoiseBandNet \\
    \midrule
Bubbles       & $35.20$  & $\mathbf{21.37}$  & $1.86$  & $\mathbf{1.15}$   & $1.2 \pm 0.3$ & $\mathbf{0.7 \pm 0.1}$  & $6.6 \pm 0.3$ & $\mathbf{4.7 \pm 0.1}$  \\
Fire          & $11.86$  & $\mathbf{2.53}$   & $6.14$  & $\mathbf{1.52}$   & $2.8 \pm 2.1$ & $\mathbf{1.7 \pm 1.0}$  & $9.6 \pm 1.3$ & $\mathbf{4.5 \pm 0.2}$  \\
Keyboard      & $13.02$  & $\mathbf{9.70}$   & $\mathbf{16.64}$ & $277.12$ & $\mathbf{5.7 \pm 2.0}$ & $20.0 \pm 7.7$ & $\mathbf{9.1 \pm 0.7}$ & $13.8 \pm 0.6$ \\
Rain          & $\mathbf{9.09}$   & $11.31$  & $\mathbf{0.98}$  & $6.19$   & $\mathbf{0.5 \pm 0.2}$ & $2.4 \pm 2.0$  & $\mathbf{9.0 \pm 0.2}$ & $9.1 \pm 0.4$  \\
River         & $\mathbf{43.66}$  & $49.85$  & $\mathbf{0.80}$  & $1.75$   & $\mathbf{0.5 \pm 0.1}$ & $0.6 \pm 0.1$  & $\mathbf{6.0 \pm 0.6}$ & $6.7 \pm 0.3$  \\
Shards        & $4.64$   & $\mathbf{1.36}$   & $\mathbf{3.79}$  & $7.58$   & $\mathbf{1.0 \pm 0.2}$ & $1.1 \pm 0.3$  & $\mathbf{7.9 \pm 0.2}$ & $8.8 \pm 0.2$  \\
Waterfall     & $\mathbf{18.23}$  & $25.88$  & $\mathbf{0.53}$  & $1.06$   & $\mathbf{0.3 \pm 0.0}$ & $0.4 \pm 0.0$  & $\mathbf{5.0 \pm 0.0}$ & $6.3 \pm 0.0$  \\
Wind          & $\mathbf{9.66}$   & $31.35$  & $\mathbf{1.95}$  & $8.48$   & $\mathbf{0.8 \pm 0.5}$ & $1.1 \pm 0.7$  & $\mathbf{5.6 \pm 0.1}$ & $5.8 \pm 0.2$ \\
    \bottomrule
    \end{tabularx}
    \caption{Validation metrics for both a \texttt{TexDSP} and a NoiseBandNet model trained on different textures. FAD metrics (lower is better) use VGGish and our proposed feature vector. Additionally, \texttt{TexStat} and MSS loss metrics are reported with means and standard deviations. In all metrics smaller is better and the best performer is highlighted in bold. \textsuperscript{(1)}Energy bands were imposed post-resynthesis. \textsuperscript{(2)}A loudness tracker was added post-resynthesis.}
    \label{tab:exp_validation}
\end{table*}

\noindent\textbf{Results:} The results highlight three key observations. First, performance varied across models, reflecting patterns observed in McDermott and Simoncelli’s work and aligning with the limitations discussed in Subsection \ref{subsec:cap_lim}. Second, although some models performed adequately, their scores remained lower than those of the reconstruction focused model NoiseBandNet. This outcome is expected, as our approach prioritizes the disentanglement of higher-level sound structures over precise reconstruction, an aspect favored by the evaluation metrics used. The latter being said, surprisingly some \texttt{TexDSP} models managed to beat its counterpart even in these metrics. Third, the metrics derived from our models appear to align more closely with our own perception of sound quality. However, to support this claim more robustly, a subjective evaluation would be necessary—an analysis that was beyond the scope of this work.\\
\noindent\textbf{Additional Comments:} A widely recognized application of the original DDSP model was timbre transfer \cite{engel2020}, where features such as pitch and loudness from an unexpected input sound are used to drive a model trained on a different instrument's timbre. For instance, applying the features of a voice recording to a violin-trained model will generate output that sounds like a violin playing the same melody. This effect is primarily due to the strong inductive bias of the model, which is trained exclusively on violin sounds and thus can only synthesize violin-like audio. Since the model operates on extracted features rather than raw audio, it naturally generates ("transfer") the timbre it was trained to pitch and loudness content extracted from the source sound, though this effect diminishes when pitch is less relevant.

In the models developed in this article, the same principle applies, though with less clear-cut results. These models retain the DDSP architecture’s bias but are designed for a wider range of sounds. For example, a fire sound can be processed by a water-trained model, resulting in a form of timbre transfer. However, unlike the pitched case, the results are harder to interpret because pitch and loudness are more meaningful for musical sounds. For texture-based sounds, meaningful timbre transfer occurs only when the input and output share a key feature from the training data. While spectral centroid and rate might seem like viable features, they lack pitch’s distinctiveness in musical contexts. Many examples of this effect are available on the article's webpage.

\section{Conclusions}\label{sec:conclusions}
This paper introduced a novel framework for advancing the analysis and synthesis of texture sounds through deep learning. Central to our contribution is \texttt{TexStat}, a loss function grounded in auditory perception and statistical modeling. By explicitly encoding key properties such as time invariance, perceptual robustness, and long-term structural focus, \texttt{TexStat} provides a formulation that is better aligned with the inherent nature of texture sounds than conventional alternatives.

In addition to \texttt{TexStat}, we presented two complementary tools: \texttt{TexEnv}, a signal processor that efficiently generates texture audio via amplitude envelope imposition, and \texttt{TexDSP}, a DDSP-based model that demonstrates the effective integration of our proposed loss into a synthesizer framework.

Our experiments validated the theoretical motivations and practical utility of \texttt{TexStat}. Specifically, when used as a feature vector, the summary statistics derived from \texttt{TexStat} outperformed general-purpose embeddings like VGGish in classification tasks. Moreover, \texttt{TexStat} exhibited improved stability against transformations such as time shifting and noise addition, providing a perceptually coherent metric for evaluating resynthesis timbre similarity.

We also demonstrated the successful application of \texttt{TexStat} as a loss function in training the \texttt{TexDSP} model, where it guided the learning process to generate indefinitely long sequences of controllable texture sounds. Although the synthesized textures differed from the input sounds, they maintained the essential perceptual qualities that define their type. Despite its strengths, we acknowledge limitations in handling pitched or rhythmically structured content, suggesting that \texttt{TexStat} is most effective when combined with other losses as a regularization component or used as an evaluation metric.

Future work should extend \texttt{TexStat} to hybrid tasks by incorporating additional loss terms for pitched and/or rhythmic sounds, and explore its applications in generative modeling, texture transformation, and neural sound design. Overall, this framework represents a promising step toward achieving better perceptual alignment in machine listening and synthesis tasks.

\section{Acknowledgments}\label{sec:acknowledgements}
This work has been supported by the project "IA y Música: Cá-tedra en Inteligencia Artificial y Música (TSI-100929-2023-1)", funded by the "Secretaría de Estado de Digitalización e Inteligencia Artificial and the Unión Europea-Next Generation EU".

\bibliographystyle{preamble/IEEEbib}
\bibliography{preamble/bibliography}


\end{document}